%% file: main.tex
\documentclass[12pt,a4paper]{scrartcl}
\usepackage[a4paper, left=1in, right=1in, top=1in, bottom=1in]{geometry}
\usepackage{setspace}
\usepackage{graphicx}
\usepackage{booktabs} 
\usepackage{amsmath, amssymb}
\usepackage{booktabs}
\usepackage{microtype}
\usepackage{hyperref}
\usepackage{authblk}
\usepackage{fvextra}
\providecommand{\keywords}[1]
{
  \small	
  \textbf{\textit{Keywords---}} #1
}

\usepackage{times}  

\usepackage[numbers]{natbib}
\usepackage{graphicx} 
\usepackage{amsmath} 

\title{\LARGE How Similar Are Grokipedia and Wikipedia?\\ {\Large A Multi-Dimensional Textual and Structural Comparison}}

\author[1,2]{Taha Yasseri$^*$}
\affil[1]{Centre for Sociology of Humans and Machines (SOHAM), Trinity College Dublin and Technological University Dublin, Dublin, Ireland}
\affil[2]{School of Mathematics and Statistics, University College Dublin, Dublin, Ireland}
\affil[ ]{$^*$Corresponding author: \texttt{taha.yasseri@tcd.ie}}

\author[1,2]{Saeedeh Mohammadi}

\date{}

\begin{document}

\maketitle
\vspace{-2cm}

\input{0-abstract}

\input{1-introduction}
\input{2-methods}
\input{3-results}
\input{4-discussion}

\section*{Acknowledgements}
This publication has emanated from research supported in part by grants from Taighde Eireann-Research Ireland under Grant numbers IRCLA/2022/3217 and 18/CRT/6049. TY acknowledges support from Workday, Inc.

\section*{Data Availability}
Data used in the study are available at \url{https://doi.org/10.5281/zenodo.19286583}.

\bibliography{main}
\clearpage
\renewcommand{\thetable}{S\arabic{table}}
\renewcommand{\thefigure}{S\arabic{figure}}

\setcounter{table}{0}  
\setcounter{figure}{0} 
\appendix

\section*{Supplementary Information for \\How Similar are Grokipedia and Wikipedia\\A Multi-Dimensional Textual and Structural Comparison\\ Taha Yasseri and Saeedeh Mohammadi\\}

\input{5-SI}

\end{document}

%% file: 0-abstract.tex
\begin{abstract}
\noindent {\Large Abstract} \\[4pt]
The launch of Grokipedia—an AI-generated encyclopedia developed by Elon Musk’s xAI—was presented as a response to perceived ideological and structural biases in Wikipedia, aiming to produce “truthful” entries using the Grok large language model. Yet whether an AI-driven alternative can escape the biases and limitations of human-edited platforms remains unclear. This study conducts a large-scale computational comparison of 17,790 matched article pairs from the 20,000 most-edited English Wikipedia pages. Using metrics spanning lexical richness, readability, reference density, structural features, and semantic similarity, we assess how closely the two platforms align in form and substance. We find that Grokipedia articles are substantially longer and contain significantly fewer references per word. Moreover, Grokipedia’s content divides into two distinct groups: one that remains semantically and stylistically aligned with Wikipedia, and another that diverges sharply. Among the dissimilar articles, we observe a systematic rightward shift in the political bias of frequently cited news media sources, concentrated primarily in entries related to history and religion, and literature and art. More broadly, the findings indicate that AI-generated encyclopedic content departs from established editorial norms, favoring narrative expansion over citation-based verification, raising questions about transparency, provenance, and the governance of knowledge in automated information systems.
\end{abstract}

%% file: 1-introduction.tex
\section{Introduction}

Online encyclopedias have become infrastructural to public knowledge, shaping how people learn about politics, science, culture, and current events. Wikipedia, launched in 2001, remains the paradigmatic example: a nonprofit, volunteer-edited project governed by policies such as Neutral Point of View (NPOV) and verifiability. Despite its success, Wikipedia has long faced questions about bias, reliability, and systemic under-representation.\footnote{For a survey of evidence and debates on reliability and bias, see e.g., \cite{GreensteinZhu2012AER, ReliabilityWikipedia2019, MIWikipediaBias2024, RightingWriters2025}.} 

In late 2025, xAI introduced Grokipedia, an AI-generated encyclopedia positioned as an explicit alternative to Wikipedia.\footnote{See \cite{GrokipediaWP, apnews2025, BI2025, Guardian2025, Verge2025, Wired2025, Time2025}.} According to xAI’s framing, Grokipedia aims to ``purge out the propaganda'' and provide ``truthful'' entries, with content generated and internally ``fact-checked'' by the Grok language model rather than curated by a community of human editors \cite{GrokipediaWP, Guardian2025, Time2025}. At launch (October 27, 2025), Grokipedia reportedly contained roughly 800-900 thousand entries, supported a suggest-edit rather than direct-edit workflow, and offered only limited transparency about licensing and code provenance \cite{GrokipediaWP, BI2025, apnews2025}. Early commentary focused on two recurring issues: (i) claims that many Grokipedia pages were copied or closely adapted from Wikipedia, and (ii) concerns about ideological bias, hallucinated citations, and other generative artifacts typical of large language models \cite{Verge2025, Wired2025, Guardian2025, Time2025}.

The launch has reignited a longstanding question: \emph{is Wikipedia biased, and if so, can an AI-first encyclopedia do better?} Empirically, prior research documents multiple bias forms in Wikipedia, early left–right slant in political pages that attenuated over time \cite{GreensteinZhu2012AER}, topical and coverage imbalances \cite{ReliabilityWikipedia2019}, and framing asymmetries across ideology and gender \cite{MIWikipediaBias2024, RightingWriters2025}. The dynamics of editorial conflict and consensus formation have also been studied extensively, showing that edit activity, controversy, and attention are strongly correlated \cite{yasseri2012dynamics}. 

Parallel research on large language models (LLMs) demonstrates measurable political and cultural biases that vary across model architectures and prompts \cite{YangMenczer2023, Rettenberger2024, Exler2025, Jenny2023}. LLMs exhibit human-like biases based on input framing \cite{lior2025wildframe} and show intrinsic political biases that affect text continuation and prediction \cite{lin2024investigating}, raising questions about the epistemic reliability of AI-generated content. Grokipedia, in particular, relies on the Grok LLM developed by xAI. In comparative studies on bibliographic reference retrieval, Grok stood out as one of the few models that did not fabricate references, although roughly 40\% of references contained incomplete or inaccurate details, such as publication year, journal name, or author information \cite{sidorenko2025assessing}. Similarly, in another study, researchers evaluated four LLMs, including Grok 2, to examine how source framing influences model judgments. Their results indicate that Grok 2 demonstrates high consistency and minimal ideological bias, with only modest shifts in evaluation when the framing of the input changed \cite{germani2025source}. These results suggest that Grok may have the potential to generate reliable and neutral encyclopedic content, although intrinsic biases in LLMs could still persist, raising the question of whether Grok can produce a less biased encyclopedia than Wikipedia.

Against this backdrop, our study asks a concrete, testable question: \emph{for titles that exist on both platforms, how similar are Grokipedia and Wikipedia in practice?} We address this by pairing articles across platforms and comparing them along three similarity dimensions: (1) \emph{lexical similarity} (TF–IDF cosine, unigram and $n$-gram overlap), (2) \emph{semantic similarity} (embedding cosine and BERTScore), and (3) \emph{stylistic similarity} (readability, lexical diversity, part-of-speech composition). In addition, we examine descriptive indicators, article length, readability, reference density, and link structure, since editorial workflows (human versus model-generated) shape article form as much as content.

We also measure the political bias of each article by aggregating the bias scores of its cited sources (see Methods for details). To contextualize these patterns, we classify all articles into topical domains and then analyse how both similarity and the shift in political bias vary across topics.

To construct a broad and representative corpus, we began with Wikipedia’s 20,000 most-edited English-language articles, a population empirically linked to controversy and social salience \cite{yasseri2012dynamics}. Among these, 17,790 titles had valid, nontrivial matches on Grokipedia.

This comparison provides empirical grounding for evaluating the distinctiveness, or sameness, of Grokipedia relative to Wikipedia. If Grokipedia largely reproduces Wikipedia’s textual and structural patterns, the platform may represent an \emph{AI-mediated continuity} rather than a substantive editorial departure. Conversely, systematic differences in length, style, or citation density could reveal an emerging form of \emph{automated editorial logic}, distinct from Wikipedia’s norms of transparency, consensus, and verifiability. By quantitatively characterizing these dimensions across matched topics, we aim to clarify where the two encyclopedias \emph{align}, where they \emph{diverge}, and what those patterns imply for neutrality, provenance, and the evolving governance of reference knowledge in the age of generative AI \footnote{For broader debates on neutrality, transparency, and provenance in AI-generated reference content, see \cite{Guardian2025, Wired2025, Time2025, yasseri2025conversation}.}.

%% file: 2-methods.tex
\section{Methods}

\subsection{Sampling and Data Collection}

We analyzed the 20,000 most-edited English-language Wikipedia articles as of November 2025, identified via cumulative edit counts. To ensure that our comparison focused on substantive articles, we excluded all list-style pages as well as titles that were date- or year-like rather than topical. Specifically, we removed (a) pages with the title “List of Topic” (e.g., List of UFC champions)  (b) calendar dates (e.g., March 11), (c) standalone years (1000–2099), (d) “Year in Topic” pages (e.g., 2003 in music), (e) “Deaths in Month Year” pages, and (f) standalone month names.

For each remaining title, we retrieved the corresponding entries from Wikipedia and Grokipedia, generating URLs of the form \texttt{https://en.wikipedia.org/wiki/<Title>} and \\ \texttt{https://Grokipedia.com/page/<Title>}.

HTML pages were downloaded between 5–11 November 2025 using the \texttt{requests} library (Python 3.11), and then parsed using \texttt{BeautifulSoup4}. After host-aware text extraction (described below), we retained only article pairs in which both platforms produced at least 500 words of clean prose. Of the original 20,000 target titles, 17,790 matched pairs met these criteria and formed the final analytical sample.

\subsubsection{Host-aware text and feature extraction}

We implemented a host-aware parsing strategy tailored to each platform’s HTML structure to maximize content fidelity.

For Wikipedia, extraction was restricted to the \texttt{\#mw-content-text} container and limited to \texttt{<p>} and \texttt{<li>} elements. Infoboxes, metadata, tables, navboxes, and reference lists were removed prior to text collection.

For Grokipedia, we used a more adaptive extractor to identify the primary article-like container (\texttt{<main>}, \texttt{<article>}, or the largest \texttt{<div>/<section>}) and retained its narrative text while filtering out menus, advertisements, and high–link-density regions. Scripts, style elements, and sidebars were removed for both platforms.

We tokenized each cleaned article into sentences and words using \texttt{nltk}'s Punkt tokenizer.

From the extracted text of each article, we computed a set of stylistic and readability metrics, in addition to lexical density and estimated reading time. All features were calculated separately for each platform.

\begin{itemize}

\item \textbf{Average sentence length:}
Computed as the mean number of tokens per sentence in the extracted text. This measure captures syntactic complexity, with longer sentences generally reflecting more complex or formally structured writing.

\item \textbf{Lexical diversity (type–token ratio):}
Defined as the number of unique word types divided by the total number of word tokens. Higher values indicate richer vocabulary usage and lower repetition within the text.

\item \textbf{Flesch–Kincaid grade level:}
A widely used readability formula that estimates the U.S. school grade level required to understand the text. It combines average sentence length and average syllables per word, with higher scores indicating more difficult material.

\item \textbf{Gunning–Fog index:}
A readability metric evaluating the number of complex words (three or more syllables) and sentence length. It approximates the years of formal education needed to comprehend a text on first reading and is commonly used to quantify formal, technical, or academic writing.

\item \textbf{Part-of-speech (POS) composition:}
Using \texttt{spaCy}'s \texttt{en\_core\_web\_sm} tagger, we computed the proportion of nouns, verbs, adjectives, and adverbs in each article. POS distribution provides a coarse-grained linguistic profile of stylistic variation, such as noun-dense factual exposition versus verb-heavy narrative text.

\item \textbf{Lexical density:}
Calculated as the ratio of unique words (types) to total words (tokens). Lexical density reflects how information-dense a text is; higher values indicate a more compact and content-rich lexical profile.

\item \textbf{Estimated reading time:}
Derived by dividing the total word count by a constant reading rate of 200 words per minute. This measure offers a standardized estimate of the amount of content a typical reader must process.

\end{itemize}

All features were computed independently for both platforms and labeled with prefixes \texttt{a\_} (Grokipedia) and \texttt{b\_} (Wikipedia).

\subsection{Similarity Measures Between Platforms}

To quantify alignment between articles on the two platforms, we computed four classes of similarity measures, each capturing a distinct dimension of textual resemblance:

\begin{enumerate}

\item \textbf{Lexical similarity:}
Lexical similarity was assessed using two complementary approaches. First, we computed cosine similarity between TF–IDF vectors constructed from 1–2 gram features with standard English stop-words removed. This measure captures similarity in word- and phrase-level vocabulary usage weighted by informativeness. Second, we calculated the unigram Jaccard index, defined as the ratio of the intersection to the union of unique word sets across platforms. This index evaluates the extent to which the two texts draw from the same vocabulary.

\item \textbf{N-gram overlap:}
To capture local phrase reuse and short-range structural similarity, we computed overlap coefficients for 1-, 2-, and 3-gram sequences. Each coefficient is defined as the size of the intersection of n-grams divided by the smaller of the two n-gram sets, providing a symmetric measure of shared phrase content independent of text length.

\item \textbf{Semantic similarity:}
We quantified semantic alignment using two embedding-based approaches. Cosine similarity between \texttt{SentenceTransformer} embeddings (\texttt{all-MiniLM-L6-v2}) measured the overall semantic closeness of the full articles in a dense representation space. In addition, we computed contextual similarity using \texttt{BERTScore}~F1 on the first 50 sentences of each article, which evaluates token-level semantic correspondence using contextual embeddings~\cite{reimers2019sentencebert}. Together, these metrics assess similarity in meaning beyond surface-level wording.

\item \textbf{Stylistic similarity:}
Stylistic similarity was evaluated by computing the Manhattan distance between the two articles’ stylistic feature vectors, which included sentence length, lexical diversity, readability metrics, and part-of-speech composition. The resulting distances were transformed linearly to a 0–1 scale, where higher values indicate stronger similarity. This measure captures convergence in writing style independent of content.

\end{enumerate}

To verify the extent to which these metrics reflected a single latent construct, we conducted a principal components analysis (PCA) on the full set of similarity measures. The first principal component accounted for the vast majority of total variance (94\%), with all measures loading positively on the component (n--gram overlap 3: 0.528, n--gram overlap 2: 0.478, BERTScore F1: 0.458, lexical Jaccard unigram: 0.377, n--gram overlap 1: 0.295, lexical TF--IDF cosine: 0.198, semantic embedding cosine: 0.103, stylistic similarity: 0.071). This indicates that the metrics are largely colinear and capture a common underlying dimension of article similarity. Based on this result, we construct a unified combined similarity score defined as the standardized value of the first principal component and use it in all subsequent analyses. The standardized similarity score has a mean of $-0.08$ and an SD of $0.63$. Figure~\ref{fig:comb_sim_dist} shows the distribution of the standardized combined similarity score. 

\subsection{Political Bias Ratings}

\begin{figure}[htbp]
\centering
\includegraphics[width=0.8\textwidth]{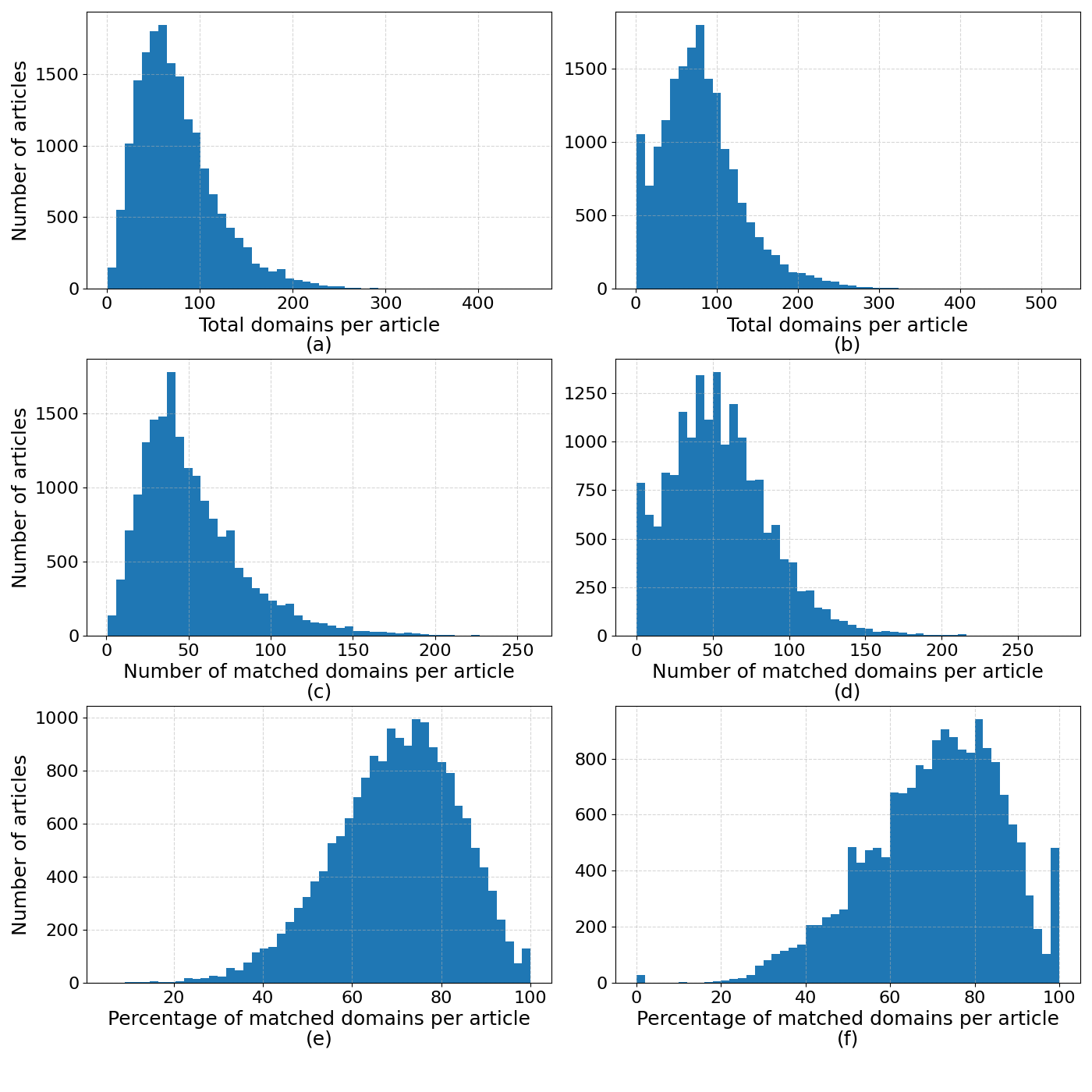}
\caption{\textbf{Distribution of matched sources across articles.} Panels (a) and (b) show the distribution of the total number of domains cited per article for Wikipedia and Grokipedia, respectively. Panels (c) and (d) show the distribution of the number of domains per article that was matched to the Yang et al dataset for Wikipedia and Grokipedia, respectively. Panels (e) and (f) report the corresponding distribution of the percentage of matched domains relative to the total number of domains cited in each article.
}
\label{fig:refs_domains_match}
\end{figure}

We extracted all outbound hyperlinks from each article and mapped their host domains to political-bias scores using a domain-level political-bias dataset. To consolidate hosts referring to the same outlet, we applied eTLD+1 normalization, supplemented by a brand key derived from the leftmost label. This procedure ensured consistent attribution across domain variants and subdomain structures, resulting in 186,678 and 219,814 unique domains for Wikipedia and Grokipedia, respectively. Bias scores correspond to the derived party leaning measure provided by Yang et al. (2025)~\cite{yang2025domaindemo}. This political bias measure relies on audience-based estimates derived from social media sharing patterns, which approximate but do not directly measure the ideological content of individual sources. Scores were assigned when a value was available for the domain or any recognized brand variant (e.g., \texttt{bbc.com} and \texttt{bbc.co.uk}). Using this dataset, approximately 17\% of unique cited domains could be matched to a political bias score (33,572 and 36,406 for Wikipedia and Grokipedia, respectively). Despite this relatively low domain-level coverage, the matched domains account for a large share of citation instances (typically 60--90\%) in both encyclopedias, reflecting their higher citation frequency relative to unmatched domains. Across the entire sample, only 29 Grokipedia articles contained no matchable references. The distribution of matched domains per article is right-skewed, with most articles containing multiple matched sources. Figure~\ref{fig:refs_domains_match} illustrates the distribution of all and matched domains per article in Wikipedia and Grokipedia.

For each article pair, the bias shift was computed as the mention-weighted mean bias of matched domains in Grokipedia minus the corresponding mean in Wikipedia (positive values indicate a rightward shift; negative values indicate a leftward shift). Because the available bias datasets primarily cover news and widely shared media domains, this measure should be interpreted as capturing differences in the political orientation of \textit{news-media-type sources}, rather than the full set of cited references.

To assess robustness, we replicated the analysis using an alternative bias dataset based on the News Media Bias and Factuality mapping by Sanchez et al. (2024) ~\cite{sanchez2024mapping}. Bias scores were converted to a symmetric numeric scale ($-1$ = left, $1$ = right). This dataset provides substantially lower domain coverage (approximately 1\% of referenced domains). Despite the large differences in coverage and dataset construction, the resulting bias patterns and overall conclusions remain consistent (See Figure~\ref{fig:bias-shift_sanchez}). 

For the pie charts in Figure~\ref{fig:bias-shift}, we restricted the analysis to the low-similarity subset (articles with a negative combined similarity score) and plotted the top 5 cumulative shares of referenced news outlets. For the pie charts, we focused exclusively on news sources, removing reference or utility hosts (e.g., \texttt{doi.org}, \texttt{archive.org}, \texttt{worldcat.org}), major platforms, and government or educational domains (\texttt{.gov}, \texttt{.edu}).

\subsection{Categorisation}

To assign each article to a topical domain, we implemented a multi-stage pipeline using a large language model (LLM). Article titles were processed in batches of up to 300 and submitted to the \textit{gpt-5-nano-2025-08-07} model with explicit instructions to assign each title to exactly one topic and to avoid the use of “Other” unless no reasonable category applied. The model returned a JSON-formatted list of titles grouped by category. Outputs with formatting errors were automatically resubmitted, with up to five retries per batch. All valid outputs were merged into an initial table mapping each title to a category. The exact prompts and API parameters are described in the SI.

Because LLM-generated labels can vary across runs, we conducted several independent classification rounds using the same model and prompt. After each round, titles labeled as “Other” were isolated and resubmitted for reclassification. The results were then merged across rounds. Titles that remained unclassified after all iterations were ultimately assigned to “Other”. This multi-round integration procedure increased coverage, reduced inconsistency, and yielded a more stable set of topic labels for all article pairs.

As a final validation step, we submitted each provisional category (batched in sets of 200 entries) to the \textit{gpt-4.1-mini} model and instructed it to flag and propose revised labels for any titles that appeared misclassified (see SI for more details). All suggested corrections were manually reviewed before inclusion in the final categorisation.

Articles were assigned to one of the following topical domains: history, politics, geography, entertainment, sports, science and technology, religion and ideologies, brands and products, languages, literature and art, health and environment, business and infrastructure, animals and nature, or other.

\subsubsection{Validation of topic classification}
To assess the reliability of the automated classification procedure described above, we conducted several validation exercises. First, the entire LLM classification pipeline was executed six times independently using the same model and prompt. Agreement across these runs was high (Fleiss’ $\kappa = 0.73$), indicating substantial consistency in the automated topic assignment.
Second, we conducted a human validation exercise. A random sample of 200 article titles was independently annotated by six human coders using the same topical categories and instructions provided to the LLM. Inter-coder agreement among the human annotators was similarly high (Fleiss’ $\kappa = 0.75$), indicating substantial agreement.
For both the automated and human annotations, a majority rule was used to determine the final topic assignment for each title. Comparing the resulting human and machine classifications yields strong agreement (Cohen’s $\kappa = 0.83$), suggesting that the automated classification procedure produces topic assignments closely aligned with human judgment.
These results indicate that the LLM-based classification is both internally stable and externally consistent with human annotations. The majority-rule aggregation of the LLM runs was used as the final topic classification in the analysis.

\subsection{Statistical and Comparative Analysis}

All analyses were performed in Python~3.11 using \texttt{pandas}, \texttt{numpy}, and \texttt{scikit-learn}. We calculated the descriptive statistics (mean~$\pm$~SD) for all features. We evaluated Differences between Grokipedia and Wikipedia using paired $t$-tests for normally distributed measures and Wilcoxon signed-rank tests otherwise. Spearman rank correlations among similarity metrics were computed to assess interdependence and clustering. Visualizations, including histograms, correlation heatmaps, and mean~$\pm$~SE summary plots, were generated with \texttt{matplotlib}.

%% file: 3-results.tex
\section{Results}

\subsection{Sample and Coverage}

From the list of the 20,000 most-edited English Wikipedia articles, we successfully retrieved and matched 17,790 Grokipedia–Wikipedia article pairs that passed all quality checks (see more details in Methods). This corpus provides substantial coverage across a wide and diverse range of topics.

\subsection{Descriptive Differences Between Platforms}

\begin{figure}[htbp]
\centering
\includegraphics[width=0.8\textwidth]{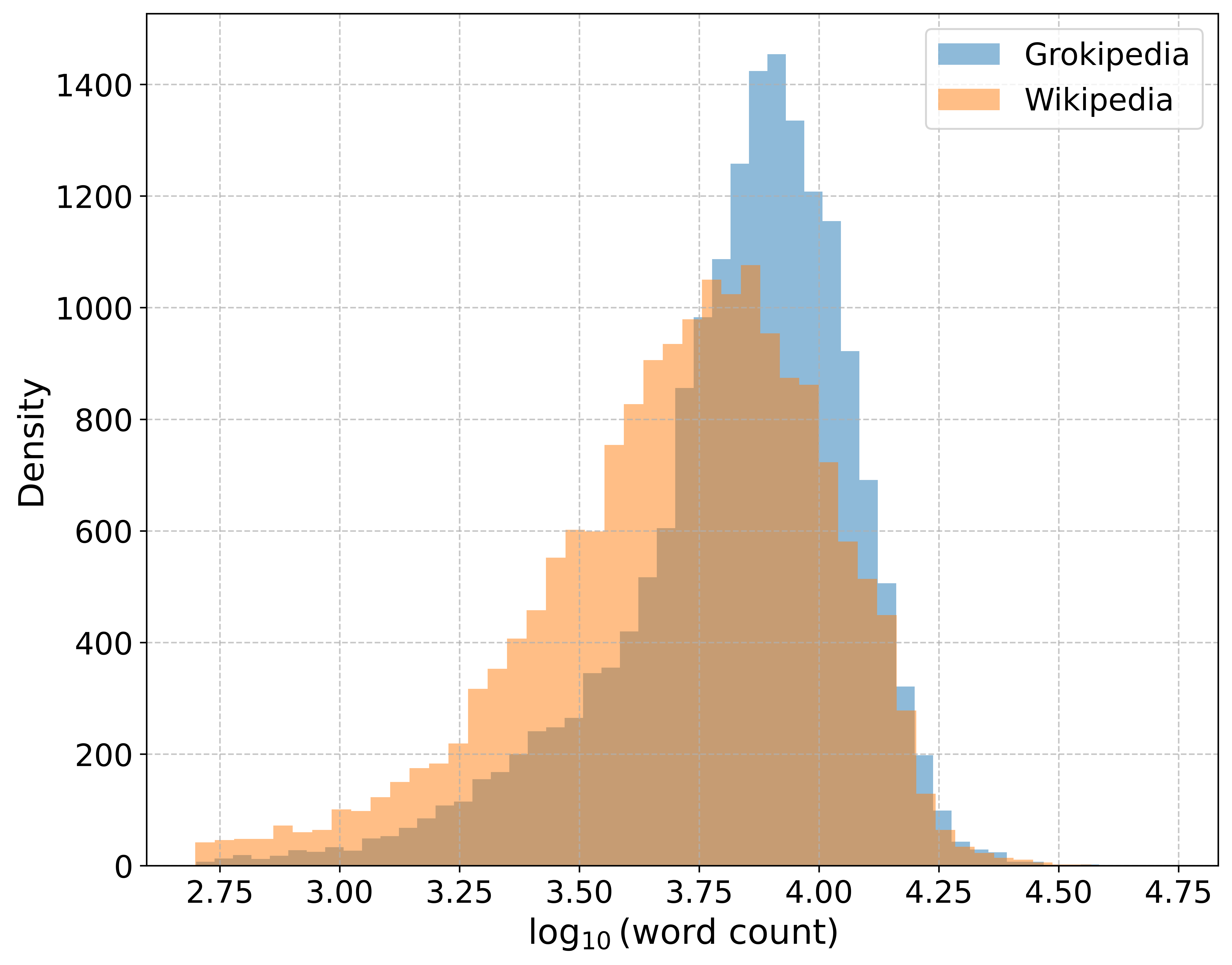}
\caption{\textbf{Distribution of article length} Distribution of $\log_{10}$ of word counts for Grokipedia and Wikipedia}
\label{fig:length}
\end{figure}

\begin{figure}[htbp]
\centering
\includegraphics[width=\textwidth]{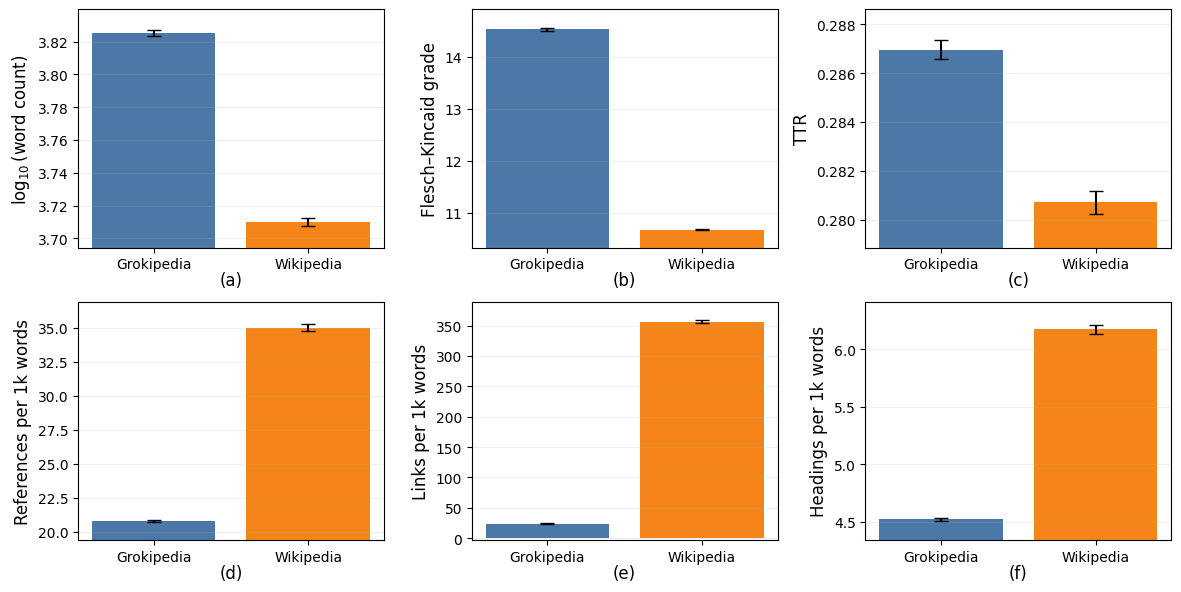}
\caption{\textbf{Platform differences for key descriptives}. 
Panels show: (a) clean word count (log$_{10}$), (b) Flesch--Kincaid grade, 
(c) lexical diversity (TTR), (d) references per 1k words, 
(e) links per 1k words, and (f) headings (H2--H4) per 1k words.
 (means $\pm$ standard error)}
\label{fig:structural_means}
\end{figure}

Figure~\ref{fig:length} compares the distribution of article lengths across platforms on a logarithmic scale. Overall, Grokipedia entries are systematically longer than their Wikipedia counterparts. While Wikipedia contains a larger number of very short articles, Grokipedia articles exhibit a pronounced peak around 7,000 words, indicating that most Grokipedia entries are substantially more verbose. Readability, measured by the Flesch--Kincaid grade level (Figure~\ref{fig:structural_means} (b)) and lexical diversity (Figure~\ref{fig:structural_means} (c), is higher on Grokipedia, indicating more syntactically complex and less reader-friendly prose.

Table~\ref{tab:structural_desc} (left panel) summarises key descriptive statistics. Across the dataset, Grokipedia articles average 7,662 words versus 6,280 on Wikipedia, and they exhibit far fewer explicit references, links, and headings per thousand words.  
These contrasts reinforce that Grokipedia expands text length through elaboration rather than citation density or structural detail. Figure~\ref{fig:structural_means} visualises these platform differences (means $\pm$~SE) across the measures reported in Table~\ref{tab:structural_desc}.

\begin{table*}[t]
\centering
\resizebox{\textwidth}{!}{
\begin{tabular}{lccccccc}
\hline
 & \multicolumn{2}{c}{All pairs (17,790)} & \multicolumn{2}{c}{Highly similar pairs (6,089)} & \multicolumn{2}{c}{Highly different pairs (11,701)} &  \\
\cline{2-7}
Metric & Wikipedia & Grokipedia & Wikipedia & Grokipedia & Wikipedia & Grokipedia & $p$-value \\
\hline
Mean article length (words) &  6280 (3828) & 7662 (3662) 
& 5062 (3296) & 5125 (3005) 
& 6915 (3932) & 8983 (3256) & $<0.001$ \\
Mean Flesch--Kincaid grade & 10.7 (1.7) & 14.5 (3.9) 
& 10.2 (1.6) & 10.0 (2.1) 
& 10.9 (1.6) & 16.9 (2.2)  & $<0.001$ \\
Mean type--token ratio &  0.28 (0.06) & 0.29 (0.05) 
& 0.28 (0.07) & 0.28 (0.07) 
& 0.28 (0.06) & 0.29 (0.04)  & $<0.001$ \\
Mean references per 1k words & 35 (37) & 20 (10) 
& 34 (44) & 21 (15) 
& 35 (33) & 21 (5)  & $<0.001$ \\
Mean hyperlinks per 1k words & 357 (323) & 24 (16) 
& 383 (375) & 30 (25) 
& 343 (292) & 21 (6)  & $<0.001$ \\
\hline
\end{tabular}}
\caption{\textbf{Descriptive differences between Grokipedia and Wikipedia articles.} Comparison of structural and linguistic properties between Wikipedia and Grokipedia articles across three groups: all article pairs (left panel), highly similar pairs (middle panel), and highly different pairs (right panel). Values represent mean (SD). The $p$-values are calculated using two-tailed paired t-tests. }
\label{tab:structural_desc}
\end{table*}

\subsection{Cross-Platform Similarity}

\begin{figure}[htbp]
\centering
\includegraphics[width=0.85\textwidth]{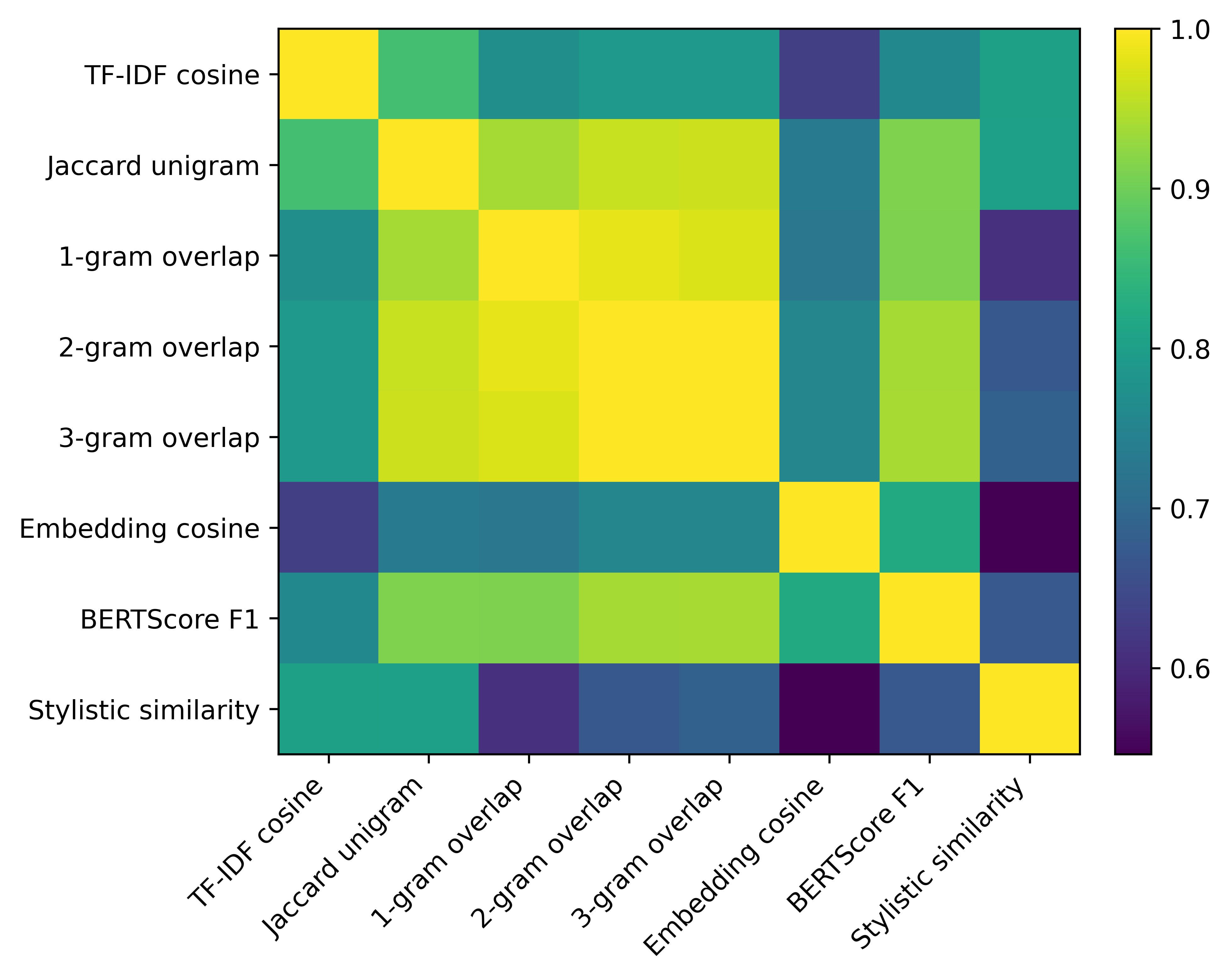}
\caption{\textbf{Correlation matrix of similarity metrics.} Correlation among similarity metrics (Pearson's $r$) across all article pairs.}
\label{fig:simheat}
\end{figure}

To assess how closely Grokipedia mirrors Wikipedia, we computed lexical, semantic, and stylistic similarity metrics for each article pair. Figure~\ref{fig:simheat} shows the correlation structure among these metrics: lexical and semantic measures form a tight cluster, while stylistic dimensions vary more independently. Table~\ref{tab:similarity} reports the average values across all pairs.

\begin{table}[htbp]
\centering
\caption{\textbf{Descriptive statistics of similarity measures.} Values represent mean (SD) for each similarity metric, along with observed minimum and maximum values. Higher scores indicate greater similarity between Grokipedia and Wikipedia articles. ($N{=}17{,}790$)}
\label{tab:similarity}
\begin{tabular}{lccc}
\toprule
Metric & Mean (SD) & Min & Max\\
\midrule
TF–IDF cosine & 0.684 (0.186) & 0.024 & 1.000\\
Jaccard (unigram) & 0.434 (0.299) & 0.046 & 1.000\\
Overlap (1-gram) & 0.640 (0.235) & 0.190 & 1.000\\
Overlap (2-gram) & 0.411 (0.373) & 0.048 & 1.000\\
Overlap (3-gram) & 0.373 (0.412) & 0.005 & 1.000\\
Semantic cosine & 0.862 (0.102) & 0.160 & 1.000\\
BERTScore (F1) & 0.306 (0.369)] & -0.606 & 0.983\\
Stylistic similarity & 0.882 (0.076) & 0.429 & 0.999\\
\bottomrule
\end{tabular}
\end{table}

\subsection{Distribution of Similarity Scores}

\begin{figure}[htbp]
\centering
\includegraphics[width=0.95\textwidth]{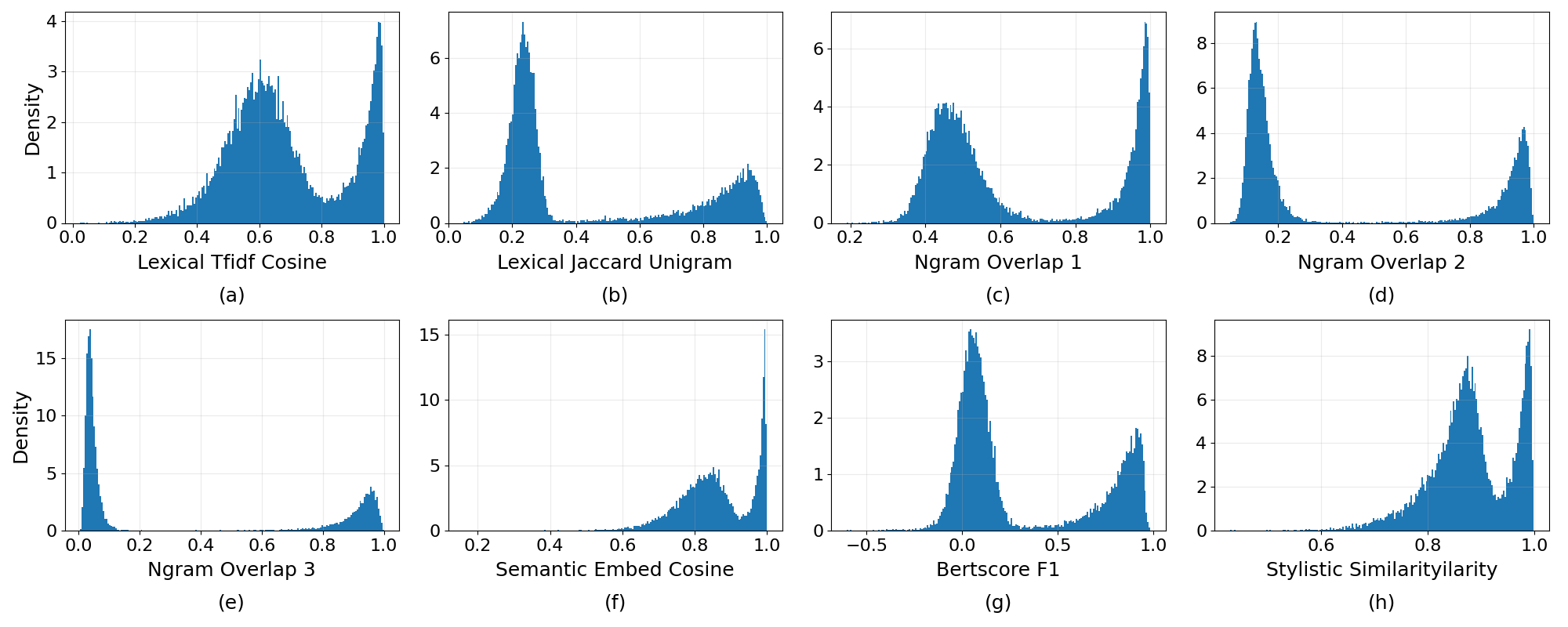}
\caption{\textbf{The distributions of similarity scores across metrics.} Each panel corresponds to one of the eight metrics. (a) TF–IDF cosine, (b) Jaccard (unigram) (c) Overlap (1-gram) (d) Overlap (2-gram) (e) Overlap (3-gram) (f) Semantic cosine (g) BERTScore (F1) (h) Stylistic similarity}
\label{fig:simdists}
\end{figure}

Figure~\ref{fig:simdists} illustrates the histograms for all eight similarity metrics. Most distributions are binomial, indicating that a portion of Grokipedia pages are moderately to highly aligned with their Wikipedia equivalents, while the others are significantly distinct.  

To reduce this multidimensional space, we conducted a principal components analysis (PCA). The first component explained the vast majority of variance, with all metrics loading in the expected direction, indicating that the measures capture a common underlying dimension of article similarity. We therefore use this component as a combined similarity score in subsequent analyses. Figure~\ref{fig:comb_sim_dist} shows the distribution of the combined similarity score.

The distribution in Figure~\ref{fig:comb_sim_dist} is distinctly bimodal, indicating two substantive groups of article pairs: one in which Grokipedia and Wikipedia differ substantially (66\% of pairs), and another in which the two versions are highly similar (34\% of pairs). As shown in Table~\ref{tab:structural_desc}, differences in descriptive measures are driven primarily by highly dissimilar pairs, in which Grokipedia articles are longer, more complex, and rely on fewer references. Among highly similar pairs, the two platforms are much closer in length, readability, and lexical diversity.

\subsection{Political Bias}
Another dimension of article change concerns shifts in political orientation. To quantify this, we examined the references cited in each article and assigned each source a bias score using publicly available datasets (see Methods for details). For each article pair, we computed the bias shift as the difference between the average bias score of the Grokipedia article and that of its Wikipedia counterpart. Positive values indicate that the Grokipedia version relies more heavily on right-leaning sources, whereas negative values reflect a shift toward left-leaning sources.

\begin{figure}[htbp]
  \centering
  \includegraphics[width=0.85\textwidth]{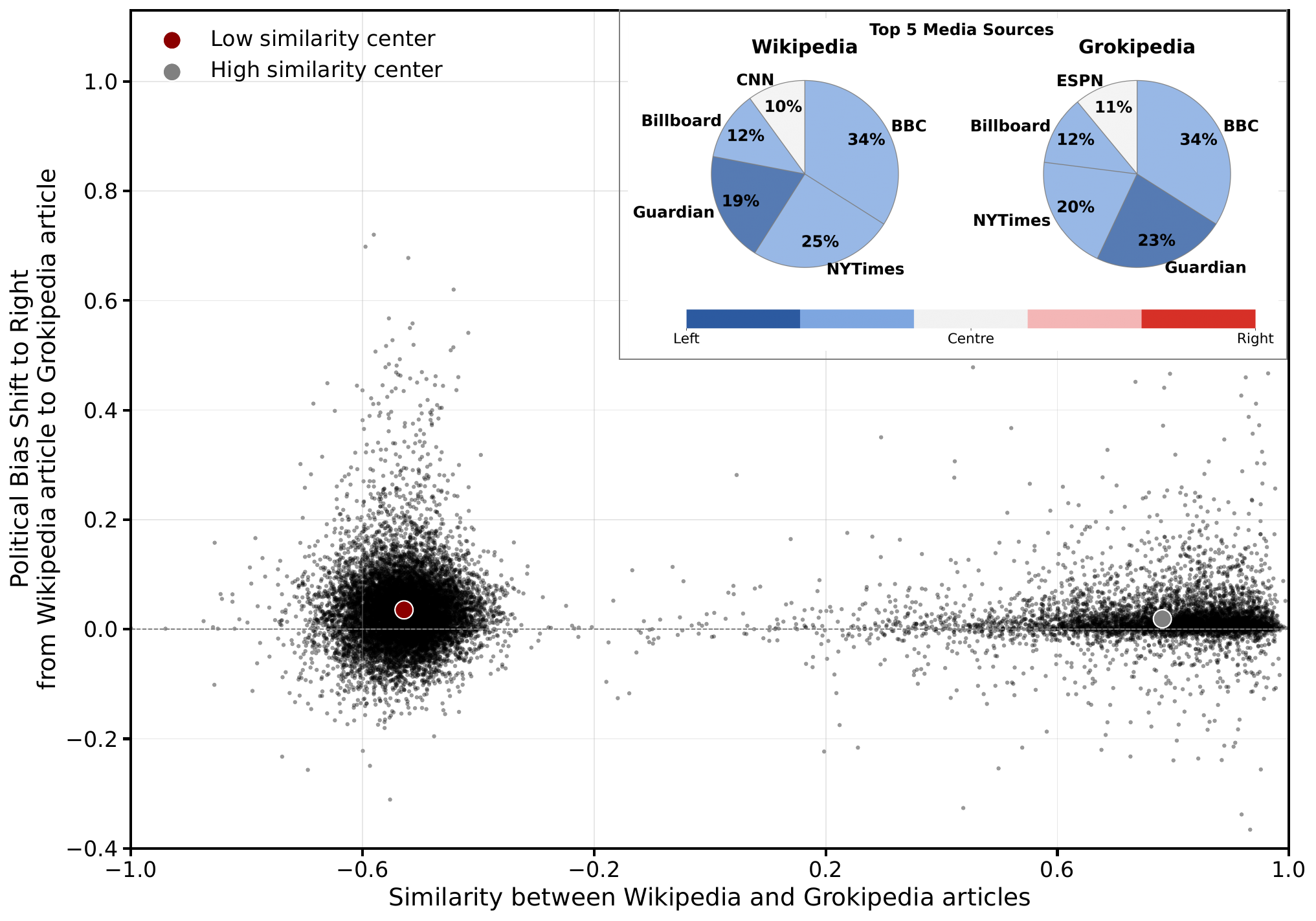}
  \caption{\textbf{Bias shift vs. article similarity.}
Each point represents an article pair. The $x$-axis shows the combined similarity score between the Wikipedia and Grokipedia versions, and the $y$-axis shows the political bias shift (Grokipedia $-$ Wikipedia). Positive values indicate a shift toward right-leaning sources, while negative values indicate a shift toward left-leaning sources. Colored markers denote the centres of mass for the low- and high-similarity groups.
\emph{Inset:} Pie charts show the top five news domains cited in Grokipedia (left) and Wikipedia (right) among articles with negative combined similarity scores, weighted by citation frequency. Slice areas represent relative contribution, and colors encode political bias.}
  \label{fig:bias-shift}
\end{figure}

Figure~\ref{fig:bias-shift} plots the bias shift for each article pair against its combined similarity score. As shown in the figure, articles that differ more substantially across platforms are markedly more likely to exhibit a right-leaning shift, whereas highly similar articles tend to cluster near zero, indicating minimal ideological movement.

The inset of Figure~\ref{fig:bias-shift} displays the five most frequently cited news sources within this low-similarity group for each platform, with wedges color-coded by ideological bias. As shown, the least similar Grokipedia articles exhibit a relative rightward shift in sourcing; however, they remain predominantly left-leaning.

\subsection{Comparison within Topics}
Because these articles cover a diverse set of subject areas, we categorized them into 14 topical domains (see Methods for details). This allows us to compare similarity and bias shift not only across individual articles but also across broader thematic domains.

\begin{figure}[htbp]
  \centering
  \includegraphics[width=0.85\textwidth]{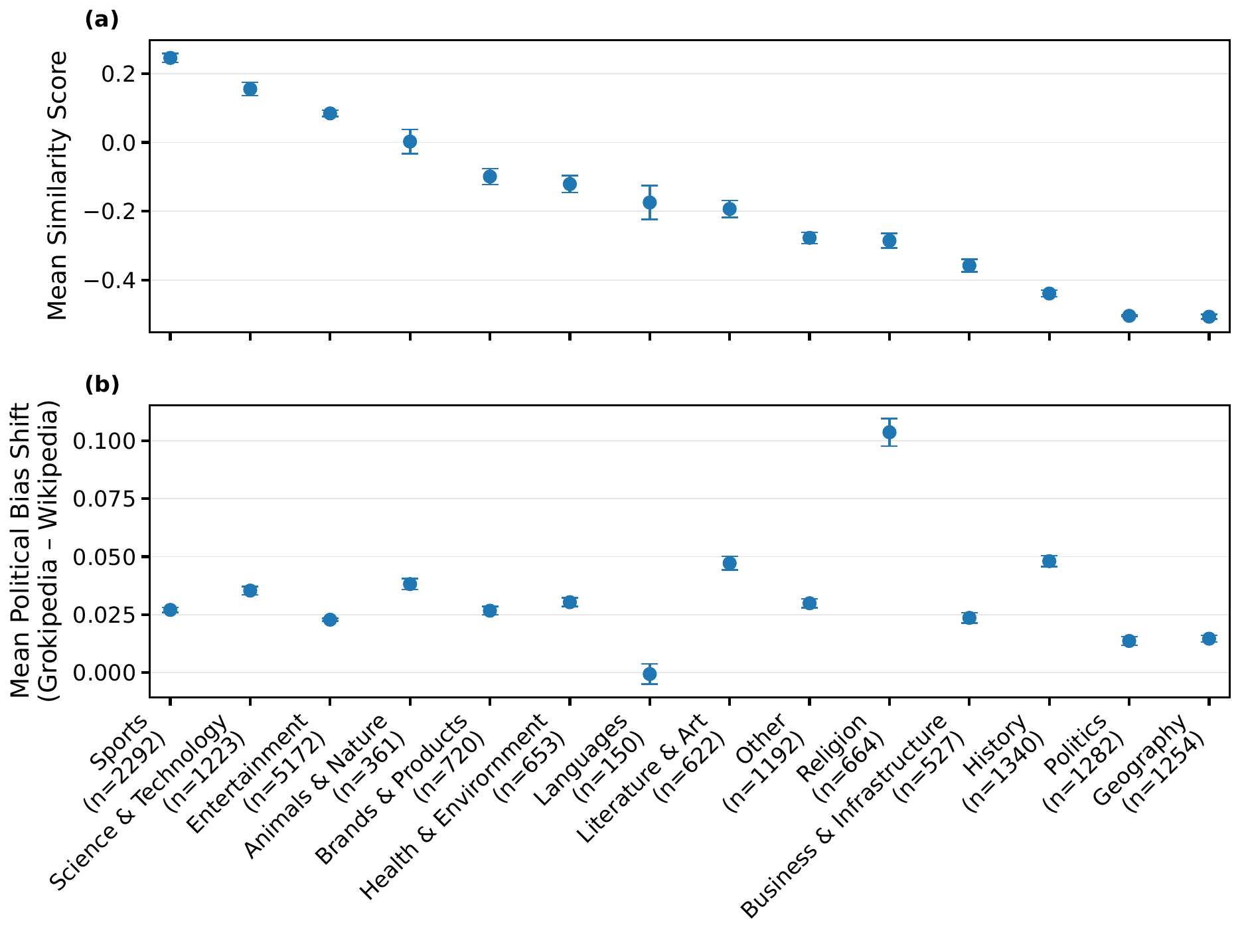}
  \caption{\textbf{Topic-level comparison of similarity and political bias shift between} Wikipedia and Grokipedia articles.
(a) Mean combined similarity scores between Wikipedia and Grokipedia articles within each topic, with standard-error bars.
(b) Mean political bias shift (Grokipedia $–$ Wikipedia) for articles in each topic, indicating whether Grokipedia content tends to lean more left or right relative to Wikipedia. Error bars show standard errors within each topic.}
  \label{fig:topic_comparison}
\end{figure}

Figure~\ref{fig:topic_comparison} summarizes these findings by plotting, for each topical category, the average similarity between Wikipedia and Grokipedia articles together with the corresponding average shift in political bias. As shown in Figure~\ref{fig:topic_comparison}(a), the largest cross-platform differences in similarity appear in articles related to geography, politics, history, business and infrastructure, and religion. In parallel, as shown in Figure~\ref{fig:topic_comparison}(b), the strongest rightward shifts in source bias occur in articles on religion, history, and literature and art.

%% file: 4-discussion.tex
\section{Discussion and Conclusion}

This study presents the first large-scale, systematic comparison between Grokipedia and Wikipedia across 17,790 matched article pairs drawn from the 20,000 most-edited English Wikipedia titles. Despite their distinct production paradigms, community-driven versus AI-generated, the two encyclopedias exhibit remarkable alignment at the level of meaning and style. Semantic similarity averages around 0.86 and stylistic similarity around 0.88, indicating that Grokipedia reproduces much of Wikipedia’s linguistic and conceptual structure.  
However, the platforms diverge substantially in form and informational scaffolding. Grokipedia articles are, on average, several times longer, with higher Flesch–Kincaid grade levels and lexical diversity (greater syntactic complexity), but reference density. This pattern suggests that Grokipedia’s generation process elaborates on existing material, expanding text length and rhetorical flow, rather than producing substantively new or more rigorously sourced knowledge.

An important distinction between the two platforms lies in their use of supporting media and internal link structure. Unlike Wikipedia, the current version of Grokipedia includes no images and does not offer linking to related pages within the platform. These omissions substantially reduce navigational affordances and make Grokipedia articles harder to scan, explore, and contextualize.

These results also resonate with earlier linguistic analyses of Wikipedia itself. Yasseri, Kornai, and Kertész~\citeyearpar{yasseri2012complexity} showed that language complexity in Wikipedia varies systematically with editorial dynamics: simplification efforts reduce syntactic depth without necessarily affecting lexical richness, and conflict intensity correlates with readability shifts.  Our findings parallel this pattern in reverse: Grokipedia’s language inflates sentence length and readability grade. In both cases, surface-level stylistic change (simplification or elaboration) alters perceived accessibility more than underlying vocabulary structure. Moreover, since our corpus also centers on heavily edited and hence often controversial topics, the observed stylistic divergence may partly reflect how both human and AI systems respond to socially contested knowledge: by expanding exposition rather than diversifying substance.

The similarity distributions reveal additional nuance. The bimodal pattern across measures indicates that while some Grokipedia pages closely mirror Wikipedia’s organisation and tone, others diverge sharply. This heterogeneity suggests a hybrid authorship logic: part replication, part improvisation. In some cases, Grok appears to reproduce Wikipedia’s structure with high fidelity; in others, it diverges, generating content that is stylistically and semantically distinct.

Elon Musk framed Grokipedia as an antidote to Wikipedia’s “propaganda” and ideological bias \citep{apnews2025, businessinsider2025, yasseri2025conversation}. Yet early independent assessments from outlets including The Verge, Wired, The Washington Post, and TIME found that many Grokipedia articles are derived from Wikipedia—often copied or paraphrased—while selectively emphasizing Musk’s personal and political narratives \citep{Wired2025, oremus2025, Verge2025, wong2025}. Our quantitative results support this view: Grokipedia articles fall into two distinct clusters, with some heavily reliant on Wikipedia pages and others exhibiting systematic divergence.

As argued by \citet{yasseri2025conversation}, Wikipedia’s openness transforms bias from a flaw into a form of \textit{epistemic visibility}: disputes, edits, and talk pages make disagreement transparent and correctable. Grokipedia inverts this model. Its authorship is singular, automated, and invisible; bias becomes latent, hidden within model weights and unseen editorial heuristics.  
This creates what might be termed an \textit{epistemic opacity paradox}: Grokipedia appears neutral because it lacks human editors, yet its underlying generative logic is uninspectable and unaccountable. The contrast highlights a deeper epistemological divide between \textit{collective knowledge systems}, where disagreement is traceable, and \textit{algorithmic knowledge systems}, where authority is inferred but not negotiated.

Complementing these textual and stylistic findings, our reference-level analysis (Figure~\ref{fig:bias-shift}) reveals that the magnitude and direction of political bias shifts between Grokipedia and Wikipedia articles are strongly related to article similarity. When Grokipedia reproduces Wikipedia closely (high similarity), the resulting entries show little or no bias shift. However, low-similarity articles display both greater variance and a rightward tendency in bias shift.  

This should be understood in the context of both methodological and substantive scope. Because available bias datasets primarily cover widely shared media domains and rely on audience-based measures derived from social media activity, our estimates capture the orientation of news media sources rather than the full spectrum of cited references or the ideological content of individual sources. At the same time, other work shows that Grokipedia and Wikipedia differ substantially in their broader composition of source types, including academic, governmental, civil-society, and user-generated sources \cite{mehdizadeh2025epistemic}. The present analysis, therefore, isolates one specific dimension of this broader epistemic shift, while other aspects of sourcing remain outside its scope.

The topic-level analysis reveals that cross-platform divergence is not evenly distributed but concentrated in specific domains. Articles on politics, geography, history, business, and religion show the lowest similarity between Grokipedia and Wikipedia, suggesting that the model’s generative behavior departs most from human-edited content precisely in areas where public narratives are contested or where factual structures are complex. Notably, some of these domains exhibit the strongest rightward shift in cited sources, too. This alignment between low similarity and increased ideological skew suggests that Grokipedia’s deviations are not merely stylistic but often substantively directional.

In effect, Grokipedia trades collective accountability for computational authority. Even though some articles do diverge from Wikipedia, they still exhibit a left-leaning bias within (albeit less than their Wikipedia counterparts). However, the fact that this rightward shift is more pronounced in specific contested articles might point to the fact that this is not the result of the LLM model generating more neutral sources but a systematic decision by the creators. This cannot be examined as the prompts and the parameters used by the model is not publicly available.

Together, these findings point to a deeper epistemological contrast. Wikipedia’s openness renders bias visible and contestable through edits, disputes, and deliberation \cite{yasseri2025conversation}. Grokipedia replaces this process with opaque, automated authorship, embedding potential bias within model behavior rather than exposing it to scrutiny. Despite its stated corrective aim, Grokipedia functions less as an epistemic alternative than as an AI-mediated reconfiguration of Wikipedia, trading collective accountability for computational authority.

Several limitations qualify these findings. First, the dataset focuses on Wikipedia’s 20,000 most-edited pages, of which 17,790 have counterparts on Grokipedia. This selection likely overrepresents high-profile, contentious topics, those most prone to editing wars or ideological scrutiny~\cite{yasseri2012dynamics}. Second, the similarity metrics employed, lexical, semantic, structural, and stylistic, quantify form and textual alignment but not factual accuracy or ideological framing. The presence of hallucinated claims, selective omissions, or subtle rhetorical shifts remains outside the scope of automated comparison. Third, the bias measure relies on the domain-level political orientation of cited sources, which is only an indirect proxy for the ideological framing of individual articles. Even with the broader bias dataset used in this study, only approximately 17\% of cited domains could be matched to a bias score, because many references in encyclopedic content point to books and documents not included in media-bias datasets. Consequently, a substantial portion of citations fall outside the measurable sample.  Fourth, both platforms are dynamic: Wikipedia continuously evolves, while Grokipedia’s generative parameters may change with future model retraining. Finally, Grokipedia’s underlying data sources and editorial interventions are opaque, preventing full provenance auditing or causal inference about model bias.

Despite these limitations, the large-scale comparative design, consistency of results across multiple metrics, and robustness checks provide a useful empirical characterization of emerging differences between AI-generated and human-edited encyclopedic knowledge systems.

Rather than functioning as a genuine epistemic alternative, Grokipedia—despite its stated aim to ``correct'' Wikipedia’s ideological slant—operates as a synthetic derivative of Wikipedia. It mirrors Wikipedia’s informational scope and linguistic tone but replaces community deliberation with algorithmic synthesis. The result is an encyclopedia that is fluent yet fragile, expansive in form but shallow in verifiability. In effect, Grokipedia trades collective accountability for computational authority.

As AI-generated reference content may itself be used to train future language models, these patterns have implications beyond a single platform. Without transparency, accountability, and sustained human oversight, such derivative systems risk propagating existing biases into the epistemic foundations of future AI systems.

Future research should move beyond surface similarity to assess factual divergence, ideological asymmetry, and user trust. Comparing how readers evaluate credibility across human-edited and AI-generated sources may reveal whether Grokipedia’s expansion of text corresponds to an expansion, or erosion, of understanding.

%% file: 5-SI.tex
\subsection*{LLM Prompt and API Parameters}
This section provides the exact prompt template and API parameters used for LLM-based topic classification. These materials are reported verbatim to ensure reproducibility.
For the first stage of the categorisation, we used the \texttt{chat.completions.create} API with the \textit{gpt-5-nano-2025-08-07} model, with the temperature set to 1. The system message and prompt used for all requests are provided below.
\hfill \break
\textbf{System message:}
\begin{Verbatim}[breaklines=true,breakanywhere=true]
You are an expert at topic classification and hierarchical clustering. Ensure the output is valid JSON. Do not include comments or trailing commas.
\end{Verbatim}
\textbf{Prompt:}
\begin{Verbatim}[breaklines=true,breakanywhere=true]
You will receive a list of titles. Your task is to classify EACH title into EXACTLY ONE of the predefined categories below.

IMPORTANT:
- Always choose the **closest** meaningful category.
- Use **"Other" only when NO reasonable category fits at all**.
- Try to avoid "Other" as much as possible.

Here are the STRICT categories you MUST choose from:
{categories_str}

Your output MUST be VALID JSON with this exact structure:

{{
  "History": [],
  "Politics_and_Political_Figures": [],
  "Geography_and_Nations_and_Cities": [],
  "Music_Movies_and_Entertainment_Celebrities": [],
  "Sports_and_Athletes_and_Teams": [],
  "Science_Technology_and_Knowledge": [],
  "Religion_and_Ideologies": [],
  "Brands_and_Products": [],
  "Languages": [],
  "Literature_and_Art": [],
  "Health_and_Environment": [],
  "Business_and_Infrastructure": [],
  "Animals_and_Nature": [],
  "Dates": [],
  "Other": []
}}

CLASSIFICATION RULES:
1. EVERY title must appear exactly once in the output.
2. DO NOT create new categories and DO NOT remove any.
3. **Avoid 'Other' unless absolutely impossible to classify elsewhere.**
4. When unsure, choose the category that is the **closest semantic match**.
5. Only use "Other" if the title genuinely does not relate to ANY listed category.
6. Output VALID JSON ONLY — no comments, no trailing commas.

Here are the titles: {batch}
\end{Verbatim}
\hfill \break
For the second stage, we used the same API function, system message, and temperature, but with the \textit{gpt-4.1-mini} model. The prompt used in this stage is provided below.

\hfill \break

\textbf{Prompt:}
\begin{Verbatim}[breaklines=true,breakanywhere=true]
You are an expert content classifier. These titles were labeled under:
**{topic}**

Allowed categories:
{allowed_topics}

For each title, check if the topic assignment is correct.
If incorrect, recommend the correct topic.

Respond only in valid JSON:
[
  {{"title": "...", "assigned_topic": "...", "correct": true/false,
    "suggested_topic": "..." }}
]

Titles:
{batch}
\end{Verbatim}
\hfill \break.
\subsection*{Additional Figures}
\begin{figure}[htbp]
\centering
\includegraphics[width=\textwidth]{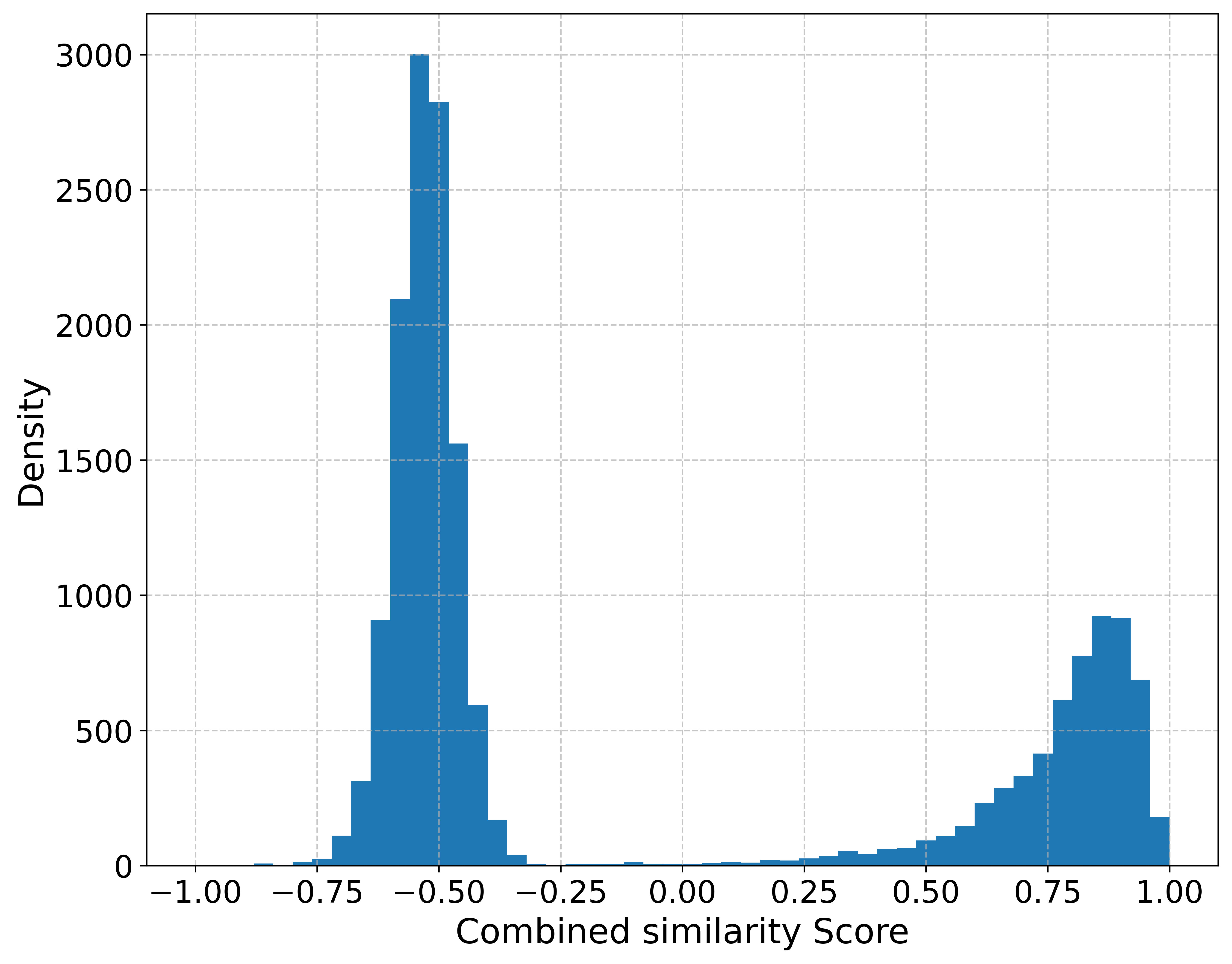}
\caption{\textbf{Combined similarity score.} Distribution of the standardized combined similarity score derived from the first principal component of the eight similarity measures.}
\label{fig:comb_sim_dist}
\end{figure}

\begin{figure}[htbp]
  \centering
  \includegraphics[width=\textwidth]{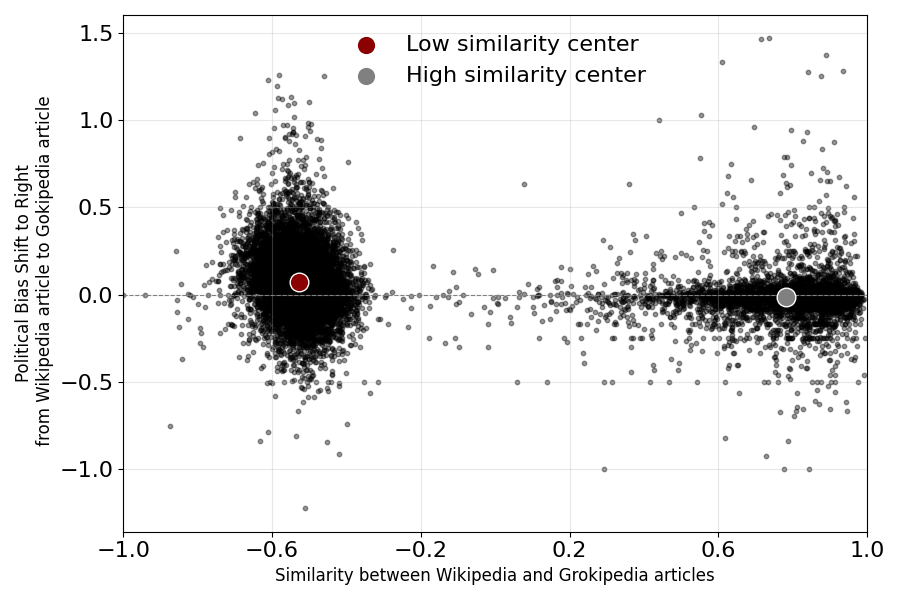}
  \caption{\textbf{Bias shift vs. article similarity.}
Each point represents an article pair. The $x$-axis shows the combined similarity score between the Wikipedia and Grokipedia versions, and the $y$-axis shows the political bias shift (Grokipedia $-$ Wikipedia), measured using the Sanchez et al.\cite{sanchez2024mapping} dataset. Positive values indicate a shift toward right-leaning sources, while negative values indicate a shift toward left-leaning sources. Colored markers denote the centres of mass for the low- and high-similarity groups. For low-similarity articles, the centre of mass shifts rightward relative to Wikipedia ($y = 0.074$), whereas high-similarity articles remain close to neutral ($y = -0.012$).}
  \label{fig:bias-shift_sanchez}
\end{figure}